\documentclass[reprint,superscriptaddress,amsmath,amssymb,pra,aps,showpacs]{revtex4-2}

\usepackage{cases}
\usepackage{amsmath,amsthm}
\usepackage {xcolor}
\usepackage{comment}

\usepackage{mathrsfs}
\usepackage{comment}
\usepackage{amssymb}
\usepackage{graphicx}
\usepackage{subfigure}

\usepackage[colorlinks,
            linkcolor=blue,
            anchorcolor=blue,
            citecolor=blue]{hyperref}

\newtheorem{theorem}{Theorem}
\newtheorem{lemma}{Lemma}

\usepackage{braket}

\usepackage{graphicx}
\usepackage{dcolumn}
\usepackage{bm}

\begin{document}

\title{Quantum state texture from the average absolutes of density entries}

\author{Cong Xu}
\email[]{2230501028@cnu.edu.cn}
\affiliation{School of Mathematics, Zhengzhou University of
Aeronautics, Zhengzhou, Henan 450046,  China}


\author{Ming-Jing Zhao}
\email[]{zhaomingjingde@126.com}
\affiliation{School of Science, Beijing Information Science and Technology University, Beijing 102206, China}

\author{Tao Li}
\affiliation{School of Mathematics and Statistics, Beijing Technology and Business University, Beijing 100048, China}

\author{Qing-Hua Zhang}
\email[]{qhzhang@csust.edu.cn}
\affiliation{School of Mathematics and Statistics, Changsha University of Science and Technology, Changsha 410114, China}

\author{Shao-Ming Fei}
\email[]{feishm@cnu.edu.cn}
\affiliation{School of Mathematical Sciences, Capital Normal University, Beijing 100048, China}

\begin{abstract}
Quantum state texture is recently recognized as an important quantum resource in quantum information tasks. The only textureless state firstly proposed in $(Phys. Rev. Lett. 133, 260801 (2024))$ is $f_1=\ket{f_1}\bra{f_1}$, where $\ket{f_1}=\frac{1}{\sqrt{d}}\sum_{i=1}^{d}\ket{i}$. We take into account the modulus of the density matrix entries, and reasonably define the textureless states to be ones that all the modulus of the density matrix entries are equal. We present a quantifier
of quantum state texture with respect to such non-convexity set of textureless states. As an application, we demonstrate in the transverse field Ising model that the texture of the subsystem of the ground state effectively characterize the quantum phase transitions, providing an alternative texture-based characterization of
quantum criticality.
We also provide the relationships among the quantum state texture, coherence and purity. Furthermore, we explore the quantum state texture based quantum speed limit in dissipative dynamical processes and derive the lower bound on the time required in texture variation for some dynamical processes.
\end{abstract}
\maketitle

\section{Introduction}

Parisio \cite{ref1} originally introduced the notion of quantum-state texture (QST) and showed that it is useful in characterizing the unknown quantum gates in universal circuit layers. Given an orthonormal basis $\mathcal{B}\equiv\{|i\rangle\}^d_{i=1}$ in a $d$ dimensional Hilbert space $\mathcal{H}$, QST can be understood as treating the density matrix $\rho$ as a three-dimensional graph. In this graph, the matrix row and column indices form the first two dimensions, while the real part of the entries $\rho_{ij}$ of $\rho$ make up the third dimension. Each quantum state corresponds to such a graph, which usually displays the irregularity or texture, a three dimensional table of columns. In this regard, the simplest table is the one with all the columns being a same number. The only existing quantum state which gives rise to such table is $f_1=\ket{f_1}\bra{f_1}$, where $\ket{f_1}=\frac{1}{\sqrt{d}}\sum_{i=1}^{d}\ket{i}$.
This state is just the textureless state given in \cite{ref1}. In \cite{ref1} the ``state rugosity" is introduced as a QST quantifier, which is proven to have potential applications in equilibrium thermodynamics. Subsequent to this pioneering research, Wang et al. \cite{Wang} explored alternative QST measures including the trace distance measure, geometric measures and fidelity-based measures, and showed that the QST measures based on fidelity can be served as a witness of an out-of-equilibrium situation.  Patra et al. \cite{Patra} explored the role of quantum state texture in probing resource theories and quantum phase transition.

We consider the general textureless states to be the ones that whose absolute values of all the density entries are the same.
In this way, the quantum state texture quantifies the uniformity of the distribution of the moduli of the density entries in the given basis. In above mentioned three-dimensional graph, the values $|\rho_{ij}|$ constitutes the third dimension, namely, the more uniformity of  $|\rho_{ij}|$, the less quantum state texture.

The remainder of this paper is organized as follows. In Sec. \uppercase\expandafter{\romannumeral2}, we introduce the quantum state texture from the average absolutes of density entries and present a texture quantifier. In Sec. \uppercase\expandafter{\romannumeral3}, using the transverse field Ising model, we demonstrate that the ground-state subsystem texture effectively characterizes quantum phase transitions. In Sec. \uppercase\expandafter{\romannumeral4}, we investigate the relationships among quantum state textures, quantum coherence and purity. We also establish the relationship between the quantum state texture of generalized Werner states (isotropic states) and the $l_1$ norm coherence.  In Sec. \uppercase\expandafter{\romannumeral5}, we investigate the quantum speed limit based on quantum state texture. The conclusion is presented in Sec. \uppercase\expandafter{\romannumeral6}.

\section{Quantum state texture based on the deviation from the average absolutes of density entries}\label{sec:2}
The only textureless (free) state $f_1=\ket{f_1}\bra{f_1}$ \cite{ref1}  is a very particular state with all density entries being $\frac{1}{d}$.
Generally, we consider textureless states to be the ones that the absolute values of all the density entries are the same.
Given an orthonormal basis $\mathcal{B}\equiv\{|i\rangle\}^d_{i=1}$ in a $d$ dimensional Hilbert space $\mathcal{H}$. A quantum state $\rho=[\rho_{ij}]$ in this basis
is textureless if $|\rho_{ij}|$ are equal for all $i,j$. Let us define the average absolutes of density entries
\begin{equation}\label{1}
A_{\mathcal{B}}(\rho)=\frac{\sum^d_{k,l=1}|\rho_{kl}|}{d^2}.
\end{equation}
Then any deviation of the density entries from $A_{\mathcal{B}}(\rho)$ implies that $\rho$ is not a textureless state. Therefore, a natural quantifier of quantum state texture corresponding to basis $\mathcal{B}=\{|i\rangle\}^d_{i=1}$ is
\begin{equation}\label{2}
T_{\mathcal{B}}(\rho)=\sum^d_{i,j=1}\left||\rho_{ij}|-A_{\mathcal{B}}(\rho)\right|,
\end{equation}
where $|\rho_{ij}|$ denotes the modulus of the element $\rho_{ij}=\langle i|\rho |j\rangle$. In general, for an arbitrary unitary operator $U$, $T_{\mathcal B}(U\rho U^{\dagger})
\neq T_{\mathcal B}(\rho)$. For the qubit state $\rho=|1\rangle\langle1|$, one has
$T_{\mathcal B}(|1\rangle\langle1|)
=\frac{3}{2}$.
Under the Hadamard transformation $H=|+\rangle\langle0|+|-\rangle\langle1|$, where
$|+\rangle=(|0\rangle+|1\rangle)/\sqrt{2}$ and $|-\rangle=(|0\rangle-|1\rangle)/\sqrt{2}$, one obtains $T_{\mathcal B}\left(H|1\rangle\langle1|H^{\dagger}\right)
=0\neq\frac{3}{2}=T_{\mathcal B}(|1\rangle\langle1|)$.

$T_{\mathcal{B}}(\rho)$ can be also rewritten as
\begin{equation}\label{SS}
\begin{aligned}
T_{\mathcal{B}}(\rho)=&\sum^d_{i,j=1}\left||\rho_{ij}|-A_{\mathcal{B}}(\rho)\right|\\
=&\sum_{|\rho_{ij}|\geq A_{\mathcal{B}}(\rho)}(|\rho_{ij}|-A_{\mathcal{B}}(\rho))+\sum_{|\rho_{ij}|< A_{\mathcal{B}}(\rho)}(A_{\mathcal{B}}(\rho)-|\rho_{ij}|)\\
=&\sum^d_{i,j=1}|\rho_{ij}|-2\sum_{|\rho_{ij}|<A_{\mathcal{B}}(\rho)}|\rho_{ij}|+(N_{-}-N_{+})A_{\mathcal{B}}(\rho)\\
=&2N_{-}A_{\mathcal{B}}(\rho)-2\sum_{|\rho_{ij}|<A_{\mathcal{B}}(\rho)}|\rho_{ij}|,
\end{aligned}
\end{equation}
where $N_{+}$ ($N_{-}$) denotes the number of terms satisfying $|\rho_{ij}|\geq A_{\mathcal{B}}(\rho)$ ($|\rho_{ij}|< A_{\mathcal{B}}(\rho)$).
It holds that $N_{+}+N_{-}= d^2$.

$T_{\mathcal{B}}(\rho)$ quantifies the uniformity of the distribution of the moduli of the density entries in the chosen basis.
In a three-dimensional graph, the row and column indices of the density matrix form the first two dimensions, while the values $|\rho_{ij}|$ constitutes the third dimension. The more uniformity of  $|\rho_{ij}|$, the less quantum state texture.

Naturally, it is an interesting question to discuss whether $T_{\mathcal{B}}(\rho)$ satisfies some useful properties in resource theory. With respect to the fixed reference basis
$\mathcal{B}$, we denote the set of
free (textureless) states by
\begin{equation*}
\mathcal F_{\mathcal B}
=\{\rho:T_{\mathcal B}(\rho)=0\}.
\end{equation*}
According to equation (\ref{2}) and $\mathrm{tr}(\rho)=1$, we obtain
$|\rho_{ij}|=\frac{1}{d}$, which implies $\mathrm{tr}\rho^2= \sum_{i,j=1}^d |\rho_{ij}|^2=1$, i.e.,
$\rho$ is a pure state. Therefore,
\begin{equation*}
\mathcal F_{\mathcal B}
=\left\{
|\psi_{\boldsymbol{\theta}}\rangle
\langle\psi_{\boldsymbol{\theta}}|
:|\psi_{\boldsymbol{\theta}}\rangle
=\frac{1}{\sqrt d}
\sum_{j=1}^{d}
e^{\sqrt{-1}\theta_j}|j\rangle
\right\},
\end{equation*}
where $\theta_1\ldots\theta_d\in[0,2\pi)$. Thus, our textureless states are precisely the maximally coherent states with respect to the chosen reference basis. Note that the set of free states is non-convex, namely, for any states $\rho_1, \rho_2\in \mathcal{F}_{\mathcal{B}}$, it is possible that $\frac{p}{2}\rho_1+\frac{(1-p)}{2}\rho_2\notin \mathcal{F}_{\mathcal{B}}$ for some $0<p<1$. It is obvious that when all the density entries are positive real numbers, our free states degenerate to the single one $f_1=\ket{f_1}\bra{f_1}$ \cite{ref1}. Accordingly,
\begin{equation*}
T_{\mathcal B}(\rho)\geq0,
~~~
T_{\mathcal B}(\rho)=0
\Longleftrightarrow
\rho\in\mathcal F_{\mathcal B},
\end{equation*}
namely, $T_{\mathcal B}(\rho)$ is faithful.

In consistent with the nonconvexity of the set of free states, the texture quantifier $T_{\mathcal B}(\rho)$ is not convex either. For example, take $\rho_{+}=|+\rangle\langle+|$ and $\rho_{-}=|-\rangle\langle-|$.
We have $T_{\mathcal B}(\rho_{+})=T_{\mathcal B}(\rho_{-})=0$.
Therefore,
\begin{equation*}
T_{\mathcal B}
\left(\frac{\rho_{+}+\rho_{-}}{2}
\right)=1>\frac{1}{2}T_{\mathcal B}(\rho_{+})
+\frac{1}{2}T_{\mathcal B}(\rho_{-}).
\end{equation*}
Furthermore, $T_{\mathcal B}$ is not additive, as obviously,
\begin{equation*}
T_{\mathcal B\otimes\mathcal B}\left(\frac{I_{2}}{2}\otimes\frac{I_{2}}{2}\right)
\neq
2T_{\mathcal B}\left(\frac{I_{2}}{2}\right).
\end{equation*}

A quantum operation defined by a set of Kraus operators ${K}_i$ satisfying $\sum^d_{i=1}{K}^{\dagger}_iK_i=I_d$ (with $I_d$ being the identity operator) is said to be a free operation if $K_i \mathcal{F}_{\mathcal{B}} K^\dagger_i \subset \mathcal{F}_{\mathcal{B}}$ for all $i$. Here, we identify an explicit class of free operations. Consider a quantum operation with a single Kraus operator $K=U_{\pi,\boldsymbol{\phi}}$ satisfying $K^{\dagger}K=I_d$ \cite{2016PRAfan},
\begin{equation*}
\Phi_{\pi,\boldsymbol{\phi}}(\rho)
=U_{\pi,\boldsymbol{\phi}}
\rho
U_{\pi,\boldsymbol{\phi}}^{\dagger},
\end{equation*}
where $U_{\pi,\boldsymbol{\phi}}=\sum_{j=1}^{d}e^{\sqrt{-1}\phi_j}
|\pi(j)\rangle\langle j|$, $\pi$ is an arbitrary permutation and
$\phi_j\in[0,2\pi)$ for $j=1,\ldots d$.
For any textureless state $|\psi_{\boldsymbol{\theta}}\rangle
\langle\psi_{\boldsymbol{\theta}}|$, we have
\begin{align*}
U_{\pi,\boldsymbol{\phi}}
|\psi_{\boldsymbol{\theta}}\rangle
&=
\frac{1}{\sqrt d}
\sum_{j=1}^{d}
e^{\sqrt{-1}(\theta_j+\phi_j)}
|\pi(j)\rangle\in
\mathcal F_{\mathcal B},
\end{align*}
namely, $U_{\pi,\boldsymbol{\phi}}
\mathcal F_{\mathcal B}
U_{\pi,\boldsymbol{\phi}}^{\dagger}
=\mathcal F_{\mathcal B}$.
By a straightforward calculation, we obtain $T_{\mathcal{B}}(\Phi_{\pi,\boldsymbol{\phi}}(\rho))=T_{\mathcal{B}}(\rho)$. Consequently, $T_{\mathcal{B}}(\rho)$ satisfies the monotonicity condition, $T_{\mathcal{B}}(\Phi_{\pi,\boldsymbol{\phi}}(\rho))\leq T_{\mathcal{B}}(\rho)$, as well as the contractivity, $|T_{\mathcal{B}}(\Phi_{\pi,\boldsymbol{\phi}}(\rho))
-T_{\mathcal{B}}(\Phi_{\pi,\boldsymbol{\phi}}(\sigma))|\leq|T_{\mathcal{B}}(\rho)-T_{\mathcal{B}}(\sigma)|$ under this class of free operations.

In the theory of quantum resources, the set of free states are not necessarily to be convex \cite{qd5,qd6}, for example, the set of Gaussian states \cite{non8,non9} and the  coherence with multi-basis.
In Refs. \cite{qd5,qd6}, the authors established the quantum resource theory for non-convex free sets by employing the generalized robustness as the quantifier of the quantum resource. However,  we cannot define the texture robustness in a similar way. Since there are no full-rank states in our set of free states, the texture robustness is equal to infinity \cite{Wang}.

In fact, our texture quantifier $T_{\mathcal{B}}(\rho)$ differs fundamentally from the texture quantifier $R(\rho)=-\ln\bra{f_1}\rho\ket{f_1}$ introduced in Ref. \cite{ref1} in the following aspects: First, the set of textureless states is different. The set of textureless states satisfying $T_{\mathcal{B}}(\rho)=0$ consists of quantum states whose element modulus are all equal, i.e., maximally coherent states. In contrast, the textureless states in Ref. \cite{ref1} consist only of the single-point $f_1=\ket{f_1}\bra{f_1}$. Clearly, our set of free states contains $f_1$. Second, the geometric interpretations are different. $T_{\mathcal{B}}(\rho)$ measures the fluctuation of the modulus of matrix elements relative to their mean, whereas $R(\rho)$ measures the negative logarithm of the projection probability of the state $\rho$ onto the uniform superposition state $f_1$. Third, in terms of quantum state discrimination, $T_{\mathcal{B}}(\rho)$ has a wider range of applicability than $R(\rho)$.  Let
\begin{equation*}
\begin{aligned}
\rho_1=
\begin{pmatrix}
0.5 & 0.25 \\
0.25 & 0.5
\end{pmatrix},
~~
\rho_2=
\begin{pmatrix}
0.5 & 0.25+0.4\sqrt{-1} \\
0.25-0.4\sqrt{-1} & 0.5
\end{pmatrix}.
\end{aligned}
\end{equation*}
One has  $R(\rho_1)=R(\rho_2)=-\ln\frac{3}{4}\approx 0.28768$, while $T_{\mathcal{B}}(\rho_1)=0.5$, $T_{\mathcal{B}}(\rho_2)\approx0.05660$.
$T_{\mathcal{B}}(\rho)$ distinguishes quantum states that share the same real parts but differ in their imaginary parts.

\section{Characterizing phase transitions based on $T_{\mathcal{B}}(\rho)$ }\label{}

The phase transitions manifest as abrupt changes in the state of a system in the thermodynamic limit \cite{QPT1,QPT2,QPT3}. At absolute zero temperature, thermal fluctuations cease to exist, and quantum fluctuations become the sole source of the corresponding phase transitions, termed as quantum phase transitions (QPTs). Various quantifiers of quantum features have been proposed to study QPTs including topological phase transitions. As applications, we show that $T_{\mathcal{B}}(\rho)$ characterizes the quantum phase transitions like
quantum coherence, quantum entanglement and quantum discord \cite{CQPT1,CQPT2,QPT5,QPT6,QPT7,QPT8,QPT9,QPT10,QPT11}.

Consider a many-body system described by the Hamiltonian $H(\lambda)$, where $\lambda$ is a tunable parameter. For each $\lambda$, let $|\Psi_g(\lambda)\rangle$ denote the ground state of the system. In the vicinity of the critical point $\lambda_c$, the ground state may undergo a dramatic change in response to an infinitesimal variation of $\lambda$. Occurring at absolute zero temperature, where thermal fluctuations are completely frozen out, a quantum phase transition is driven purely by quantum fluctuations induced by the tuning of $\lambda$.

Denote $\sigma^{1}$, $\sigma^{2}$, $\sigma^{3}$ and $\sigma^{0}=I$ the standard Pauli matrices and identity. We consider the transverse field Ising model (TFIM) with the Hamiltonian given by
\begin{equation}\label{}
\begin{aligned}
H_{I}=-J\sum_{j=1}\sigma^{3}_j\otimes\sigma^{3}_{j+1}-B\sum_{j=1}\sigma^{1}_j.
\end{aligned}
\end{equation}
We choose the free parameter $\lambda:=\frac{J}{B}$ to explore the phase diagram. The model can be solved analytically using the Jordan-Wigner transformation \cite{34,35,36,CQPT2}. The one-site reduced density operator for the ground state is given by the expectation values of the Pauli matrices,
\begin{equation}\label{22}
\begin{aligned}
\rho_\ast=\frac{\langle\sigma^0\rangle\sigma^0+\langle\sigma^1\rangle\sigma^1
+\langle\sigma^2\rangle\sigma^2+\langle\sigma^3\rangle\sigma^3}{2}.
\end{aligned}
\end{equation}
 Here $\langle\cdot\rangle$ denotes the ground-state expectation value. The critical point is located at the second-order phase transition point $\lambda_c=1$.  The system is gapped for $0<\lambda<1$, where the ground state resides in the paramagnetic phase with a vanishing order parameter $\langle \sigma^{3}\rangle=0$. For $\lambda>1$, the system is also gapped but exhibits a double degeneracy in the energy spectrum.

In the thermodynamic limit, the symmetry can be spontaneously broken, and the (degenerate) ground states reside in the ferromagnetic phase characterized by a nonvanishing order parameter $\langle \sigma^3 \rangle$. The two branches of degenerate ground states differ by the opposite sign of the order parameter, labeled $|0_{+}\rangle$ and $|0_{-}\rangle$, respectively. In this phase, the system may also reside in an equal mixture of $|0_{+}\rangle$ and $|0_{-}\rangle$, i.e., $\rho_{\mathrm{TG}} = (|0_{+}\rangle \langle 0_{+}| + |0_{-}\rangle \langle 0_{-}|)/2$, which exhibits a vanishing order parameter. This mixed state is referred to as the ``thermal ground state.''

We examine the texture for the thermal ground state.
According to Refs. \cite{8,CQPT2}, for the thermal ground state $\rho_{\mathrm{TG}}$, one has
\begin{equation}\label{23}
\begin{aligned}
\langle \sigma^1 \rangle = \frac{1}{\pi} \int_0^\pi  \frac{1 + \lambda \cos \phi}{\sqrt{1 + \lambda^2 + 2\lambda \cos \phi}} d\phi,
\end{aligned}
\end{equation}
and $\langle \sigma^2 \rangle=\langle \sigma^3 \rangle=0$. Substituting these results into equation (\ref{22}), we get
\begin{equation}\label{224}
\begin{aligned}
\rho^{TG}_\ast=
\begin{pmatrix}
 \frac{1}{2} & \frac{\langle \sigma^1 \rangle}{2} \\
\frac{\langle \sigma^1 \rangle}{2} & \frac{1}{2}
\end{pmatrix}.
\end{aligned}
\end{equation}
From equation (\ref{2}) we have

\begin{equation}\label{25}
\begin{aligned}
T_{\mathcal{B}}(\lambda)
&=2\left|\frac{1}{2}-\frac{1+|b(\lambda)|}{4}\right|
+2\left|\frac{|b(\lambda)|}{2}-\frac{1+|b(\lambda)|}{4}\right|\\
&=1-|b(\lambda)|=1-b(\lambda)\\
&=1-\frac{1}{\pi} \int_0^\pi  \frac{1 + \lambda \cos \phi}{\sqrt{1 +\lambda^2 + 2\lambda \cos \phi}} d\phi,
\end{aligned}
\end{equation}
where $T_{\mathcal{B}}(\lambda)=T_{\mathcal{B}}(\rho^{TG}_\ast)$ and $b(\lambda)=\langle \sigma^1 \rangle$.
Note that for $\lambda>1$, the integrand appearing in the definition of $b(\lambda)$ is not pointwise nonnegative. Nevertheless, its integral remains positive. For $0\leq\lambda<1$,
this follows immediately from
$1+\lambda\cos\phi\geq1-\lambda>0$. For $\lambda>1$, let
$c=\cos\phi\in[0,1]$ for $\phi\in[0,\pi/2]$, and define
\begin{equation*}
f_{\lambda}(\phi)
=\frac{1+\lambda\cos\phi}
{\sqrt{1+\lambda^{2}+2\lambda\cos\phi}}.
\end{equation*}
Then
\begin{align*}
f_{\lambda}(\phi)+f_{\lambda}(\pi-\phi)
&=\frac{1+\lambda c}{\sqrt{1+\lambda^{2}+2\lambda c}}
+\frac{1-\lambda c}{\sqrt{1+\lambda^{2}-2\lambda c}}.
\end{align*}
When $\lambda c\leq1$, both terms on the right-hand side are
nonnegative. When $\lambda c>1$, we compare the magnitude of the
negative second term with the first one. By direct calculation, it follows
\begin{align*}
&
\frac{(1+\lambda c)^2}
{1+\lambda^{2}+2\lambda c}
-\frac{(\lambda c-1)^2}
{1+\lambda^{2}-2\lambda c}
\nonumber\\
&\qquad =
\frac{4\lambda^{3}c(1-c^{2})}
{\left(1+\lambda^{2}+2\lambda c\right)
 \left(1+\lambda^{2}-2\lambda c\right)}
\geq0.
\end{align*}
Hence
$f_{\lambda}(\phi)+f_{\lambda}(\pi-\phi)\geq0$. Consequently,
\begin{equation*}
b(\lambda)
=\frac{1}{\pi}
\int_{0}^{\pi/2}
\left[f_{\lambda}(\phi)+f_{\lambda}(\pi-\phi)
\right]d\phi\geq0
\end{equation*}
for every finite $\lambda\geq0$.

We define the texture susceptibility as
\begin{equation}\label{26}
\chi_{T} \equiv \frac{\partial T_{\mathcal{B}}(\lambda)}{\partial \lambda}=-\frac{db(\lambda)}{d\lambda}=\frac{\lambda}{\pi} \int_0^\pi \frac{\sin^2\phi}{\left(1 + \lambda^2
+ 2\lambda \cos\phi\right)^{3/2}} \, d\phi .
\end{equation}
We employ the texture susceptibility to characterize the non-analyticity of the ground state at the phase transition point.
From equations (\ref{25}) and (\ref{26}), we have Figure 1. From the blue solid line in Figure 1, we see a clear change in the vicinity of the transition points $\lambda_c=1$, which demonstrates that the texture susceptibility $\chi_T$ of the thermal ground state faithfully detects the critical points.
\begin{figure}[tbp]
  \centering
  \includegraphics[width=8.6cm]{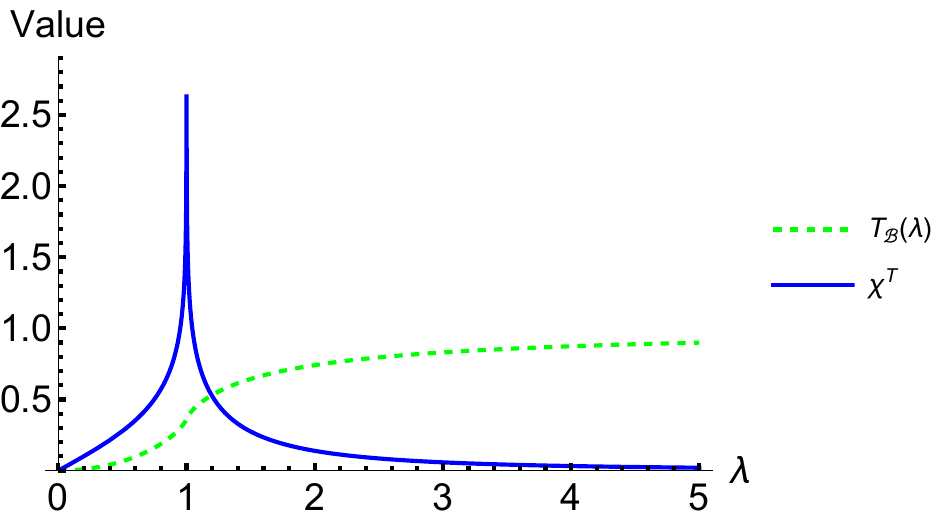}\\
\caption{Ground state of the Ising model. The green dashed line and the blue solid line represent the texture of the one-site reduced density matrix of the thermal ground state $\rho_{TG}$ and its texture susceptibility, respectively.
\label{fig:Fig1}}
\end{figure}

In Ref.\cite{CQPT2}, the non-analyticity of the ground state at the phase transition point is characterized by the singularity of the relative entropy of coherence susceptibility, $\chi_{re} \equiv \frac{\partial C_{re}(\rho)}{\partial \lambda}$, where $C_{re}(\rho)$ is the relative entropy of coherence \cite{2014prl}.
For quantum state $\rho^{TG}_\ast$, we have
\begin{equation*}
C_{re}(\rho^{TG}_\ast)=S((\rho^{TG}_\ast)_{diag})-S(\rho^{TG}_\ast)
=1-H_2\left(\frac{1+b(\lambda)}{2}\right).
\end{equation*}
Here, $S(\cdot)$ denotes the von Neumann entropy,
$(\rho^{TG}_\ast)_{diag}$ is the state obtained by removing all off-diagonal elements of $\rho^{TG}_\ast$, and $H_2(x)=-x\log_2 x-(1-x)\log_2(1-x)$ is the binary entropy. Its susceptibility is
\begin{align}\label{XRE}
\chi_{re}
&\equiv
\frac{\partial C_{re}(\lambda)}{\partial\lambda}=
\frac{1}{2}\log_2\left(\frac{1+b(\lambda)}{1-b(\lambda)}\right)
\frac{db(\lambda)}{d\lambda},
\end{align}
where $C_{re}(\lambda)=C_{ re}(\rho^{TG}_\ast)$.

Combining equations (\ref{26}) and (\ref{XRE}), we also observe that
\begin{equation}\label{RETT}
\chi_{re} =-\frac{1}{2}\log_2\left(\frac{1+b(\lambda)}{1-b(\lambda)}\right)
\chi_T.
\end{equation}
Since $b(1)=\frac{2}{\pi}$ at the critical point $\lambda_c=1$, both $\chi_{T}$ and $\chi_{re}$ exhibit singularities at this critical point. Equation (\ref{RETT}) shows that the newly introduced texture quantifier provides an alternative characterization of the same ground-state non-analyticity in the above model.

In fact, for the one-site state $\rho^{TG}_\ast$ considered above,
\begin{equation*}
T_{\mathcal{B}}(\rho^{TG}_\ast) =1-|b(\lambda)|=1-C_{l_1}(\lambda)=1-b(\lambda),
\end{equation*}
where $C_{l_1}(\lambda)=C_{l_1}(\rho^{TG}_\ast)=|b(\lambda)|=b(\lambda)$ is the $l_1$ norm coherence. Therefore,
$\chi_{T} =-\chi_{l_1}$,
where $\chi_{l_1}=\frac{dC_{l_1}(\lambda)}{d\lambda}$ is the $l_1$ norm coherence susceptibility. Therefore, for this particular one-parameter family of one-site reduced states, the texture susceptibility and the coherence susceptibility contain exactly the same information about the location and singular behavior of the quantum critical point, up to an overall sign. This example demonstrates that the state-texture framework provides an alternative and complementary
characterization of the same critical behavior from the perspective of the non-uniformity of the moduli of density-matrix
entries.

\section{Quantum state texture $T_{\mathcal{B}}(\rho)$, coherence and purity}\label{sec:2}

Quantum coherence \cite{2014prl}, arising from the superposition principle, has been investigated both theoretically and experimentally \cite{Coherence1,Coherence3,Coherence4}. We explore the relationship between the texture quantifier $T_{\mathcal{B}}(\rho)$ and quantum coherence.

\begin{theorem}
For any incoherent state, $T_{\mathcal{B}}(\rho)$ satisfies the following inequality,
\begin{equation}\label{}
T_{\mathcal{B}}(\rho)\geq2\left(1-\frac{1}{d}\right).
\end{equation}
\end{theorem}

\begin{proof}
For any incoherent state $\rho=\sum^{d}_{i=1}p_i\ket{i}\bra{i}$, with $p_i\geq0$, $\sum^{d}_{i=1}p_i=1$ and $\sum^{d}_{i,j=1}|p_{ij}|=1$,
we have
\begin{equation*}
\begin{aligned}
T_{\mathcal{B}}(\rho)=&\sum^d_{i,j=1}\left||\rho_{ij}|-\frac{1}{d^2}\right|
=\sum^d_{i=1}\left|p_{i}-\frac{1}{d^2}\right|+1-\frac{1}{d}\\
\geq&\left|\sum^d_{i=1}(p_{i}-\frac{1}{d^2})\right|+1-\frac{1}{d}
=2(1-\frac{1}{d}).
\end{aligned}
\end{equation*}
\end{proof}

Specifically, when all $ p_i\geq \frac{1}{d^2}$ and $\sum^{d}_{i=1}p_i=1$, the equation $T_{\mathcal{B}}(\rho)=2(1-\frac{1}{d})$ holds. Therefore, if $T_{\mathcal{B}}(\rho)<2\left(1-\frac{1}{d}\right)$, \(\rho\) must be coherent. This condition provides a simple sufficient criterion of coherence.

The widely used $l_1$ norm coherence measure \cite{2014prl} is defined by
\begin{equation}\label{cl1}
C_{l_1}(\rho)=\sum_{i\neq j}|\rho_{ij}|=\sum_{i,j}|\rho_{ij}|-1.
\end{equation}
We explore the relationship between $C_{l_1}(\rho) $ and $T_{\mathcal{B}}(\rho)$ based on the general Werner states and generalized isotropic states \cite{LiN}. Consider the general Werner states,
\begin{equation}\label{}
\sigma=\frac{d-\theta}{d^3-d}I_d\otimes I_d+\frac{d\theta-1}{d^3-d}F, \,\, \theta\in[-1,1]
\end{equation}
where $F=\sum^d_{a,b=1}\ket{a}\bra{b}\otimes\ket{b}\bra{a}$ with $\{\ket{a}:a=1,2,\ldots,d \}$ being an orthonormal basis of the $d$-dimensional Hilbert space and $I_d$ the corresponding identity operator. We have the following conclusion,  see the proof in Appendix A:
\small{
\begin{equation*}
\begin{split}
&T_{\mathcal{B}}(\sigma) = \\
&\begin{cases}
\dfrac{2(d^2-1)}{d^2}-\dfrac{2}{d^2}C_{l_1}(\sigma), & C_{l_1}(\sigma)\in[0, c_1) \\[2pt]
\dfrac{2(d^3-2d+1)}{d^3}(1+C_{l_1}(\sigma)), & C_{l_1}(\sigma)\in[c_1, c_3] \\[2pt]
(2-\frac{2}{d}-\frac{4}{d^2}+\frac{4}{d^3})+(2+\frac{2}{d}-\frac{4}{d^2}+\frac{4}{d^3})C_{l_1}(\sigma),   &C_{l_1}(\sigma)\in(c_3, 1]
\end{cases}
\end{split}
\end{equation*}}
where $C_{l_1}(\sigma)=\frac{|d\theta-1|}{d+1}$ is the $l_1$ norm coherence, $c_1=\frac{d-1}{d^3-d+1}$ and $c_3=\frac{d^2-1}{d^2+1}$.

For the general isotropic states,
\begin{equation}\label{}
\tau=\frac{1-\vartheta}{d^2-1}I_d\otimes I_d+\frac{d^2\vartheta-1}{d^2-1}\ket{\Psi}\bra{\Psi}, \,\,  \vartheta\in[0,1]
\end{equation}
where $\ket{\Psi}=\frac{1}{\sqrt{d}}\sum^d_{a=1}\ket{aa}$, the following equation holds, see proof in Appendix B,
\begin{equation*}
\begin{split}
T_{\mathcal{B}}(\tau)=
&\begin{cases}
\frac{2(d^2-1)}{d^2}-\frac{2}{d^2}C_{l_1}(\tau), & C_{l_1}(\tau)\in [0, c_1) \\[10pt]
\frac{2(d^3-2d+1)}{d^3}(1+C_{l_1}(\tau)),  & C_{l_1}(\tau)\in [c_1, c_2]  \\[10pt]
\frac{2(d-1)}{d^2}+\frac{2(d^2+d-1)}{d^2}C_{l_1}(\tau),  & C_{l_1}(\tau)\in (c_2, d-1]
\end{cases}
\end{split}
\end{equation*}
where $C_{l_1}(\tau)=\frac{|d^2\vartheta-1|}{d+1}$ is the $l_1$ norm coherence, $c_2=\frac{(d^2-1)(d-1)}{d^2+d-1}$.

Interestingly, when the generalized Werner states and the generalized isotropic states reduce to the Werner states \cite{W} and the isotropic states \cite{I}, respectively, the relationship between $T_{\mathcal{B}}(\rho)$ and the $l_1$ norm coherence $C_{l_1}(\rho)$ is described by the same piecewise linear function,
\begin{equation*}
\begin{split}
T_{\mathcal{B}}(\rho_{WI})=
&\begin{cases}
\frac{3}{2}-\frac{C_{l_1}(\rho_{WI})}{2}, & C_{l_1}(\rho_{WI})\in [0, \frac{1}{7}) \\[10pt]
\frac{5}{4}(1+C_{l_1}(\rho_{WI})),  & C_{l_1}(\rho_{WI})\in[\frac{1}{7},\frac{3}{5}]  \\[10pt]
\frac{1}{2}+\frac{5}{2}C_{l_1}(\rho_{WI}),  & C_{l_1}(\rho_{WI})\in(\frac{3}{5},1]
\end{cases}
\end{split}
\end{equation*}
where $C_{l_1}(\rho_{WI})$ denote the $l_1$ norm coherence of either the Werner states or the isotropic states.

Figure 2 shows that for both the Werner states and the isotropic states, $T_{\mathcal{B}}(\rho_{WI})$ is monotonically decreasing when $0\leq C_{l_1}(\rho_{WI})<\frac{1}{7}$, and reaches its minimum value of $\frac{10}{7}$ at
$C_{l_1}(\rho_{WI})=\frac{1}{7}$. For $\frac{1}{7}\leq C_{l_1}(\rho_{WI})\leq\frac{3}{5}$, $T_{\mathcal{B}}(\rho_{WI})$ is monotonically increasing, and continues to increase monotonically when $\frac{3}{5}<C_{l_1}(\rho_{WI})\leq 1$, reaching its maximum value of $3$ at $C_{l_1}(\rho_{WI})=1$.
\begin{figure}[tbp]
  \centering
  \includegraphics[width=8.6cm]{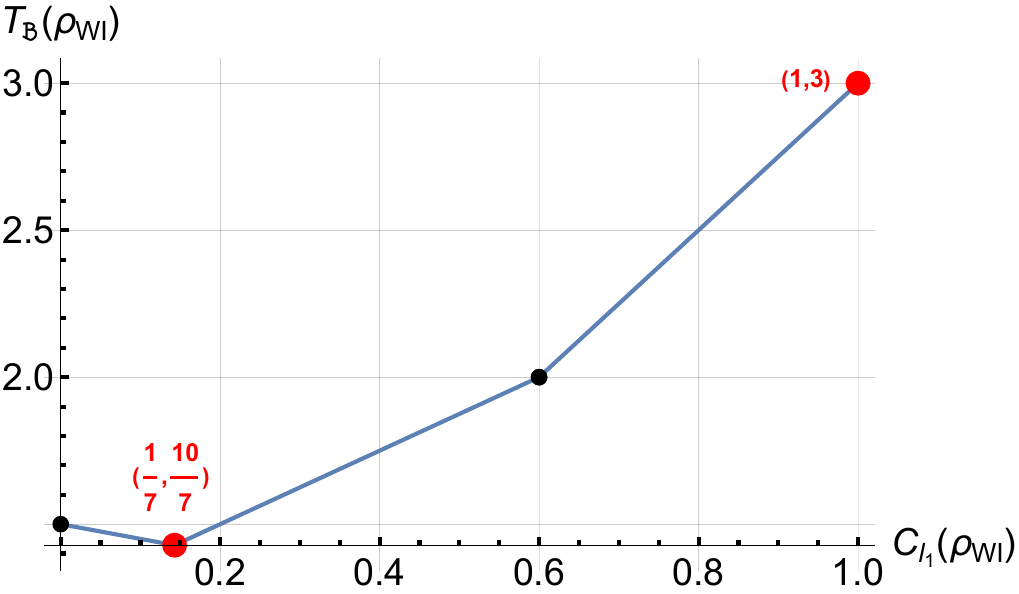}\\
\caption{ The horizontal axis represents the $l_1$ norm coherence $C_{l_1}(\rho_{WI})$ of the Werner states or isotropic states, the vertical axis represents the quantum state texture $T_{\mathcal{B}}(\rho_{WI})$. \label{fig:Fig1}}
\end{figure}

Concerning the relationship among $T_{\mathcal{B}}(\rho)$, $C_{l_1}(\rho)$ and the purity $P(\rho)=\mathrm{tr}\rho^2= \sum_{k,l=1}^d |\rho_{kl}|^2$ for a general quantum state $\rho$, we have, see the proof in Appendix C:

\begin{theorem}\label{TH2}
For a general quantum state $\rho$, the $l_1$ norm coherence $C_{l_1}(\rho)$, the purity $P(\rho)$ and the texture quantifier $T_{\mathcal{B}}(\rho)$ satisfy the following inequality,
\begin{equation}\label{FFF}
 C_{l_1}(\rho)+T_{\mathcal{B}}(\rho)\leq d\sqrt{2P(\rho)}-1.
\end{equation}
\end{theorem}

Note that for arbitrary $d$ and $P(\rho)$, the upper bound is generally not tight. In particular, the equality can  not be  attained for odd $d$ and $P(\rho)<\frac{1}{2}$. For every even $d$, consider
\begin{equation*}
\rho_{\mathrm{eq}} = \frac{1}{2}|\psi_1\rangle\langle\psi_1| + \frac{1}{2}|\psi_2\rangle\langle\psi_2|,
\end{equation*}
where
\begin{equation*}
|\psi_1\rangle = \sqrt{\frac{2}{d}}\sum_{j=1}^{d/2}|j\rangle,~~~
|\psi_2\rangle = \sqrt{\frac{2}{d}}\sum_{j=d/2+1}^{d}|j\rangle.
\end{equation*}
By direct computation, we have
\begin{equation*}
P(\rho_{\mathrm{eq}}) = \frac{1}{2},~~~  C_{l_1}(\rho_{\mathrm{eq}}) = \frac{d}{2} - 1, ~~~ T_{\mathcal{B}}(\rho_{\mathrm{eq}}) = \frac{d}{2}.
\end{equation*}
Hence,
\begin{equation}\label{DDD}
T_{\mathcal{B}}(\rho_{\mathrm{eq}}) + C_{l_1}(\rho_{\mathrm{eq}}) = d\sqrt{2P(\rho_{\mathrm{eq}})} - 1.
\end{equation}

For qubit systems, a single-qubit state can be written as
\begin{equation}\label{P}
\begin{aligned}
\rho_B=\frac{1}{2}(I_2+ \overrightarrow{r} \cdot \overrightarrow{\sigma})=\frac{1}{2}
\begin{pmatrix}
1+r_z & r_x-\mathrm{i}r_y \\
r_x+\mathrm{i}r_y & 1-r_z
\end{pmatrix},
\end{aligned}
\end{equation}
where $\overrightarrow{\sigma}=(\sigma^1,\sigma^2,\sigma^3)$ and $\overrightarrow{r}=(r_x,r_y,r_z)$ is the Bloch vector satisfying
$|\overrightarrow{r}| \leq1$, with equality for pure states. A straightforward calculation yields
\begin{align}\label{AAA}
T_\mathcal{B}(\rho_B)&=\frac{1}{4}\left(|1-r_{\perp}+2r_z|
+|1-r_{\perp}-2r_z|+2(1-r_{\perp})\right)
\nonumber\\
&=\frac{1}{2}\left[1-r_{\perp}+\max\left\{1-r_{\perp},2|r_z|\right\}
\right],
\end{align}
\begin{equation}\label{BBB}
\begin{aligned}
C_{l_1}(\rho_B)= \sqrt{r^2_x+r^2_y}\equiv r_\bot,
\end{aligned}
\end{equation}
and
\begin{equation}\label{CCC}
\begin{aligned}
P(\rho_B)=\mathrm{tr}\rho_B^2=\frac{1+r^2_x+r^2_y+r^2_z}{2}.
\end{aligned}
\end{equation}
Figure 3 illustrates the functional curves of the two sides of equation (\ref{FFF}) in the qubit case, and Figure 4 illustrates their difference (diff). It is observed from Figures 3 and 4 that Theorem 2 always holds, with the equality saturated by incoherent states, which agrees with equation (\ref{DDD}).
\begin{figure}[tbp]
  \centering
  \includegraphics[width=8.7cm]{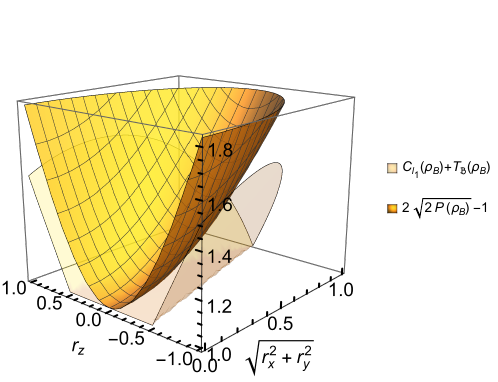}\\
\caption{The horizontal axes represent $r_z$ and $\sqrt{r^2_x+r^2_y}$, the vertical axis represents the values. The yellow meshed surface corresponds to the right-hand side of equation (\ref{FFF}), whereas the unmeshed surface corresponds to the left-hand side of equation (\ref{FFF}).}
\end{figure}
\begin{figure}[tbp]
  \centering
  \includegraphics[width=8.4cm]{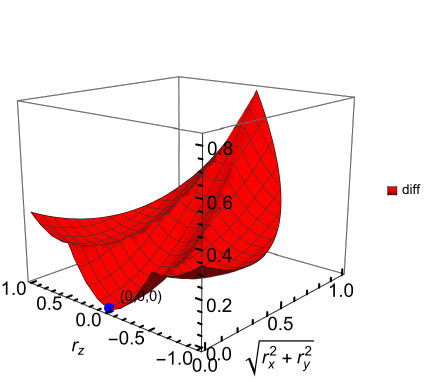}\\
\caption{The horizontal axes represent $r_z$ and $\sqrt{r^2_x+r^2_y}$. The vertical axis represents the values. The red meshed surface represents the difference (diff) between the right-hand side and the left-hand side of equation (\ref{FFF}). The blue point at (0, 0, 0) indicates the incoherent state, where the equality in equation (\ref{FFF}) is attained.}
\end{figure}

Equation (\ref{AAA}) can also be rewritten as
\begin{equation*}
T_\mathcal{B}(\rho_B)=
\begin{cases}
1-r_{\perp},
&
2|r_z|\le1-r_{\perp},
\\[2mm]
\dfrac{1-r_{\perp}}{2}+|r_z|,
&
2|r_z|\ge1-r_{\perp}.
\end{cases}
\end{equation*}
In particular, $T(\rho_B)\ge |r_z|$.
Therefore, we have the following relation,
\begin{align*}
2P(\rho_B)
&=
1+r_{\perp}^{2}+r_z^{2}
\nonumber\\
&=
1+C_{l_1}^{2}(\rho_B)+r_z^{2}
\nonumber\\
&\le
1+C_{l_1}^{2}(\rho_B)+T_{\mathcal{B}}^{2}(\rho_B),
\end{align*}
where the equality is saturated at $r_{\perp}=1$ and $r_z=0$, i.e., for the textureless state.

\section{Quantum speed limit based on quantum state texture}\label{sec:3}

The quantum speed limits (QSLs) \cite{MT} specify the minimum time required for a quantum system to evolve from a given initial state to a target state, establishing a fundamental lower bound on the evolution time \cite{MT,ML,speed3,speed4} and holding significant importance in such as quantum metrology \cite{met}, quantum control \cite{con} and quantum thermodynamics \cite{therm1,therm2}. In recent years, researchers have conducted many studies on the quantum speed limits of resource variation, including coherence \cite{CHO1,CHO2,CHO3}, entanglement \cite{entanglement1,entanglement2,entanglement3} and imaginary \cite{imaginary1}. We emphasize that the general strategy of bounding a finite variation of a quantity by an integrated dynamical speed is not specific to
quantum state texture. The purpose of the present section is instead to establish a texture-specific speed limit for the quantity $T_{\mathcal B}(\rho)$ introduced in this work.

In particular, our construction differs from the relative-entropy resource speed limit \cite{Campaioli2022} given by $M(\rho) = \min_{\sigma\in \mathbb{F}} S(\rho\|\sigma)$, where $S(\rho\|\sigma)=\mathrm{tr}(\rho\log \rho)-\mathrm{tr}(\rho \log \sigma)$ denotes the quantum relative entropy, and $\mathbb{F}$ represents the set of free states in the resource theory. However, in our texture framework, free states $\sigma \in \mathcal{F}_{\mathcal{B}}$ are rank-one maximally coherent states. Consequently, for any non-textureless state $\rho$, since $\operatorname{supp}(\rho) \nsubseteq \operatorname{supp}(\sigma)$, we have $S(\rho\Vert\sigma)=+\infty$. Therefore, this relative-entropy approach cannot provide a well-defined  texture quantifier for generic states.

Consider the continuous dynamical evolution of \(\rho_{t}\) over the time interval $t\in[0,T]$. The following theorem provides a lower bound on the time required in texture variation for some dynamical processes, see proof in Appendix D.


\begin{theorem}\label{TH3}
Let $t\mapsto\rho_t$, $t\in[0,T]$, be an absolutely continuous
trajectory of density operators on a finite-dimensional Hilbert
space. With respect to the fixed reference basis $\mathcal B$, define
\begin{equation}
    \overline{v}_{\mathcal B}
    =
    \frac{1}{T}
    \int_0^T
    \sum_{i,j}
    |\frac{d}{dt}\rho_{ij}(t)|\,dt.
\end{equation}
If $\overline{v}_{\mathcal B}>0$, the evolution time $T$ required
for the initial state $\rho_0$ to reach $\rho_T$ satisfies the following inequality,
\begin{equation}
    T
    \geq
    T_{\mathrm{TSL}}
    :=
    \frac{
    |T_{\mathcal B}(\rho_T)-T_{\mathcal B}(\rho_0)|
    }{
    2\overline{v}_{\mathcal B}
    }.
    \label{eq:texture_speed_limit}
\end{equation}
If $\overline{v}_{\mathcal B}=0$, then $\rho_t=\rho_0$ for all
$t\in[0,T]$ and consequently
$T_{\mathcal B}(\rho_T)=T_{\mathcal B}(\rho_0)$.
\end{theorem}

Theorem 3 applies to quantum dynamics whose density operator trajectory $t\mapsto\rho_t$ is absolutely continuous on the finite time interval $[0,T]$ in a finite-dimensional Hilbert space. This condition ensures the applicability of the fundamental theorem of calculus to each matrix element $\rho_{ij}(t)$, rendering the time-averaged derivative $\overline{v}_{\mathcal B}$ well-defined. In particular, this assumption encompasses continuous and piecewise-continuous unitary and Lindblad dynamical processes.

The lower bound obtained here is also different from coherence speed limits given in such as Ref. \cite{CHO1}. Those results impose constraints on the variation of coherence measures, whereas our result imposes a constraint on the variation of
$T_{\mathcal B}(\rho)$.  The distinction is structural: the free states of the resource theory of coherence are incoherent states,  whereas the textureless states considered here are maximally coherent states.  Furthermore, the texture speed limit derived below comes directly from a global Lipschitz property of
$T_{\mathcal B}$ and involves only the entrywise evolution speed
$\sum_{i,j}|\frac{d}{dt}\rho_{ij}(t)|$.  Consequently, in contrast to coherence speed limits, it requires neither an optimization over free states nor a matrix logarithm. Our bound provides a complementary bound tailored to the dynamical variation of quantum state texture.


For a detailed dissipative process, let us consider a two-level atom with the ground state $\ket{1}\bra{1}$ and the excited state $\ket{0}\bra{0}$ interacting with a dissipative environment. The time evolution of the atom is governed by the Lindblad master equation \cite{CHO1,NJP2}, which is expressed as follows:
\begin{equation*}\label{}
\begin{aligned}
\frac{d\rho_t}{dt}=\frac{\gamma}{2}(\sigma_-\rho_t\sigma_+-\frac{1}{2}\{\sigma_+\sigma_-,\rho_t\}),
\end{aligned}
\end{equation*}
where  \(\gamma\) represents the dissipation rate, $\sigma_+=\ket{0}\bra{1}$ and $\sigma_-=\ket{1}\bra{0}$ are the raising and lowering operators, respectively.

If the atom is initially in the state $\rho_0=\ket{\psi(0)}\bra{\psi(0)}$, where $\ket{\psi(0)}=\cos\frac{\theta}{2}\ket{0}+\sin\frac{\theta}{2}\ket{1}$, then the solution to the Lindblad equation can be expressed as
\begin{equation}\label{19}
\begin{aligned}
\rho_t=
\begin{pmatrix}
e^{-\frac{\gamma t}{2}}\cos^2\frac{\theta}{2} & e^{-\frac{\gamma t}{4}}\sin\frac{\theta}{2}\cos\frac{\theta}{2} \\
e^{-\frac{\gamma t}{4}}\sin\frac{\theta}{2}\cos\frac{\theta}{2} & 1-e^{-\frac{\gamma t}{2}}\cos^2\frac{\theta}{2}
\end{pmatrix}.
\end{aligned}
\end{equation}
Calculating the quantum state texture in the computational basis $\{|0\rangle , |1\rangle\}$, we have the specific expression of $T_{\mathrm{TSL}}$ given in Appendix E. As shown in Figure 5, we plot $T_{\mathrm{TSL}}$ as a function of $T \in (0, 5]$, with $\gamma=2$ and $\theta \in \{\frac{\pi}{2}, \frac{2\pi}{3}, \frac{7\pi}{8}\}$. From Figure 5, we observe that the initial state with identical matrix entries ($\theta = \frac{\pi}{2}$) exhibits a longer speed limit time for texturing, compared with the cases of $\theta=\frac{2\pi}{3}$ and $\theta = \frac{7\pi}{8}$. A textureless state exhibits a larger speed limit time than the states with nonzero initial texture.
\begin{figure}[tbp]
  \centering
  \includegraphics[width=8.6cm]{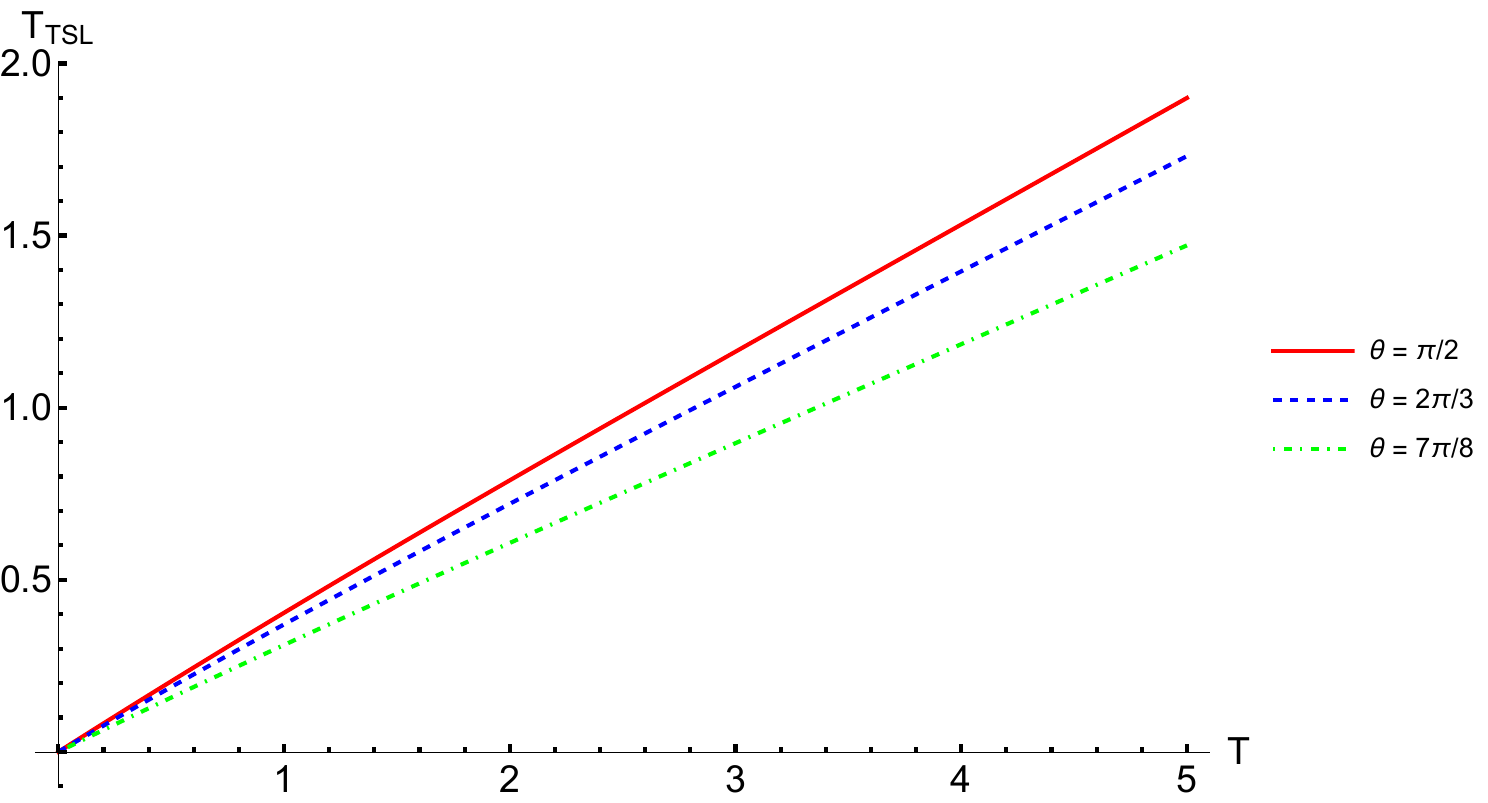}\\
\caption{$T_{TSL}$ vs $T$ for $\gamma=2$, and $\theta\in\{\frac{\pi}{2},\frac{2\pi}{3},\frac{7\pi}{8}\}$. \label{fig:Fig3}}
\end{figure}

\section{Conclusion}
We have taken into account the modulus of the entries of density matrices to define the quantum state texture and provide corresponding texture quantifiers. The textureless state in reference \cite{ref1} is only a particular free state of our set of free states.
As applications, we have shown that the texture susceptibility exactly detects the quantum phase transitions in the transverse field Ising model. Moreover, we have explored the relationship between the textures of generalized Werner states and generalized isotropic states and their $l_1$ norm coherence. We have also found that the textures of Werner states and isotropic states share the same expression with their $l_1$ norm coherence.
We also have investigated the relationships among the quantum state texture, coherence and purity.
Finally, we have also studied the quantum speed limit based on quantum state textures, providing a lower bound on the time required for texture variation in some dynamical process.

In conclusion, our texture quantifier
$T_{\mathcal{B}}(\rho)$ possesses direct dynamical significance in the quantum speed-limit relation derived in this work, where the variation $|T_{\mathcal{B}}(\rho_{T})-T_{\mathcal{B}}(\rho_{0})|$ provides a lower bound on the time required for a dynamical process to generate or erase quantum state texture. Furthermore, the texture susceptibility serves as a physically useful indicator of quantum criticality in the transverse-field Ising model. The complete resource theory of quantum state texture, including the characterization of the maximal class of free operations, the operational interpretation of $T_{\mathcal{B}}(\rho)$ in terms of resource conversion, distillation etc., remains an interesting direction for future investigation.

\section{APPENDIX}\label{sec7}

\subsection{Texture of the general Werner states}
The general Werner states $\sigma$ can be rewritten as:
\begin{equation*}\label{}
\sigma=\frac{d-\theta}{d^3-d}\sum^d_{a,b=1}\ket{ab}\bra{ab}
+\frac{d\theta-1}{d^3-d}\sum^d_{a,b=1}\ket{ab}\bra{ba}.
\end{equation*}
In the computational basis $\{\ket{ab}\}^{d}_{a,b=1}$, the matrix elements are:
 \begin{equation*}\label{}
\sigma_{ab,a'b'}=A\delta_{aa'}\delta_{bb'}
+B\delta_{ab'}\delta_{ba'},
\end{equation*}
where $A=\frac{d-\theta}{d^3-d}$, $B=\frac{d\theta-1}{d^3-d}$.
Let $k=d\theta-1\in[-d-1,d-1]$, then
$A=\frac{d^2-k-1}{d^2(d^2-1)}$, $B=\frac{k}{d(d^2-1)}$.

Type I: When $a=a'$, $b=b'$ and $a=b$ , \(\sigma\) has $n_1=d$ diagonal entries $\sigma_{aa,aa}=A+B$.

Type II: When $a=a'$, $b=b'$ but $a\neq b$, \(\sigma\) has $n_2=d^2-d$ diagonal entries $\sigma_{ab,ab}=A$.

Type III: When $a=b'$, $b=a'$ but $a\neq a'$ , \(\sigma\) has  $n_3=d^2-d$ off diagonal entries $\sigma_{ab,ba}=B$. The remaining $n_0=d^4-2d^2+d$ matrix entries are zero.

We denote $v_1=|A+B|$, $v_2=|A|$ and $v_3=|B|$. Then the $l_1$ norm coherence measure $C_{l_1}(\sigma)=|n_3v_3|=\frac{|k|}{d+1}$, the total sum of the matrix entry modules is  $n_1|A+B|+n_2|A|+n_3|B|=1+C_{l_1}(\sigma)$, and the average value is $A_{\mathcal{B}}(\sigma)=\frac{1+C_{l_1}(\sigma)}{d^4}$.

Next, we discuss two cases separately: $k\geq0$ and $k<0$.

Case A: $k\geq0$, that is, $\theta\geq\frac{1}{d}$. In this case, $k=d\theta-1\in[0,d-1]$, $C_{l_1}(\sigma)=\frac{k}{d+1}\in[0,\frac{d-1}{d+1}]$, $k=(d+1)C_{l_1}(\sigma)$. Then
$A=\frac{d^2-k-1}{d^2(d^2-1)}=\frac{1}{d^2}-\frac{C_{l_1}(\sigma)}{d^2(d-1)}$, $B=\frac{k}{d(d^2-1)}=\frac{C_{l_1}(\sigma)}{d(d-1)}$ and
$A+B=\frac{1+C_{l_1}(\sigma)}{d^2}$, which are all positive. Hence, $v_1=\frac{1+C_{l_1}(\sigma)}{d^2}$, $v_2=\frac{1}{d^2}-\frac{C_{l_1}(\sigma)}{d^2(d-1)}$ and $v_3=\frac{C_{l_1}(\sigma)}{d(d-1)}$.

We obtain $v_1-A_{\mathcal{B}}(\sigma)=(1+C_{l_1}(\sigma))\frac{d^2-1}{d^4}>0$, $v_2> A_{\mathcal{B}}(\sigma)$ if and only if $C_{l_1}(\sigma)<\frac{(d^2-1)(d-1)}{d^2+d-1}$. Denote $c_2=\frac{(d^2-1)(d-1)}{d^2+d-1}$. It follows that when $C_{l_1}(\sigma)< c_2$, $v_2> A_{\mathcal{B}}(\sigma)$; conversely, when $C_{l_1}(\sigma)\geq c_2$, $v_2\leq A_{\mathcal{B}}(\sigma)$.
Similarly, $v_3\geq A_{\mathcal{B}}(\sigma)$ if and only if $C_{l_1}(\sigma)\geq\frac{d-1}{d^3-d+1}$. Denote $c_1=\frac{d-1}{d^3-d+1}$. Accordingly, when $C_{l_1}(\sigma)\geq c_1$, $v_3\geq A_{\mathcal{B}}(\sigma)$, whereas when $C_{l_1}(\sigma)< c_1$, $v_3< A_{\mathcal{B}}(\sigma)$. It can be derived that $c_1<\frac{d-1}{d+1}<c_2$. Therefore, for Case A, the range of $C_{l_1}(\sigma)$ can be divided into the following two intervals:

Interval $(\mathrm{i})$: $0\leq C_{l_1}(\sigma)< c_1$. In this interval $v_1>A_{\mathcal{B}}(\sigma)$, $v_2> A_{\mathcal{B}}(\sigma)$ and $v_3< A_{\mathcal{B}}(\sigma)$;

Interval $(\mathrm{ii})$: $c_1\leq C_{l_1}(\sigma)\leq \frac{d-1}{d+1}$. In this interval, $v_1>A_{\mathcal{B}}(\sigma)$, $v_2> A_{\mathcal{B}}(\sigma)$ and $v_3\geq A_{\mathcal{B}}(\sigma)$.

According to equation (\ref{SS}), for Interval $(\mathrm{i})$,
$$
T_{\mathcal{B}}(\sigma)=2(n_0+n_3)A_{\mathcal{B}}(\sigma)-2n_3v_3=\frac{2(d^2-1)}{d^2}-\frac{2C_{l_1}(\sigma)}{d^2}.
$$
For Interval $(\mathrm{ii})$,
$$
T_{\mathcal{B}}(\sigma)=2n_0A_{\mathcal{B}}(\sigma)=\frac{2(d^3-2d+1)}{d^3}(1+C_{l_1}(\sigma)).
$$

Case B: $k<0$, that is, $\theta<\frac{1}{d}$. In this case, $k=d\theta-1\in[-d-1,0)$, $C_{l_1}(\sigma)=\frac{-k}{d+1}\in(0,1]$, $k=-(d+1)C_{l_1}(\sigma)$. Then $A+B=\frac{1-C_{l_1}(\sigma)}{d^2}\geq0$,
$A=\frac{d^2-k-1}{d^2(d^2-1)}=\frac{1}{d^2}+\frac{C_{l_1}(\sigma)}{d^2(d-1)}$ and $B=\frac{k}{d(d^2-1)}=-\frac{C_{l_1}(\sigma)}{d(d-1)}$. That is,
$v_1=\frac{1-C_{l_1}(\sigma)}{d^2}$, $v_2=\frac{1}{d^2}+\frac{C_{l_1}(\sigma)}{d^2(d-1)}$ and $v_3=\frac{C_{l_1}(\sigma)}{d(d-1)}$.

Therefore,  $v_1\geq A_{\mathcal{B}}(\sigma)$ if and only if $C_{l_1}(\sigma)\leq\frac{d^2-1}{d^2+1}$. Denote $c_3=\frac{d^2-1}{d^2+1}$. It follows that when $C_{l_1}(\sigma)\leq c_3$, $v_1\geq A_{\mathcal{B}}(\sigma)$; conversely, when $C_{l_1}(\sigma)>c_3$, $v_1<A_{\mathcal{B}}(\sigma)$. $v_2-A_{\mathcal{B}}(\sigma)=\frac{d^2-1}{d^4}+\frac{d^2-d+1}{d^4(d-1)}C_{l_1}(\sigma)>0$. As for $v_3$, it is the same as Case A.

It can be derived that $c_1<c_3<1$. Therefore, under Case B, the range of $C_{l_1}(\sigma)$ can be divided into following three intervals:

Interval I: $0\leq C_{l_1}(\sigma)< c_1$: $v_1>A_{\mathcal{B}}(\sigma)$, $v_2>A_{\mathcal{B}}(\sigma)$ and $v_3<A_{\mathcal{B}}(\sigma)$;

Interval II: $c_1\leq C_{l_1}(\sigma)\leq c_3$: $v_1\geq A_{\mathcal{B}}(\sigma)$, $v_2>A_{\mathcal{B}}(\sigma)$ and $v_3\geq A_{\mathcal{B}}(\sigma)$;

Interval III: $c_3< C_{l_1}(\sigma)\leq 1$: $v_1<A_{\mathcal{B}}(\sigma)$, $v_2>A_{\mathcal{B}}(\sigma)$ and $v_3>A_{\mathcal{B}}(\sigma)$.

According to equation (\ref{SS}), for Interval I and Interval II, $T_{\mathcal{B}}(\sigma)$ is the same as intervals ($\mathrm{i}$) and ($\mathrm{ii}$) in case A, respectively. For Interval III,
\begin{equation*}\label{}
\begin{aligned}
T_{\mathcal{B}}(\sigma)&=2(n_0+n_1)A_{\mathcal{B}}(\sigma)-2n_1v_1\\
&=(2-\frac{2}{d}-\frac{4}{d^2}+\frac{4}{d^3})+(2+\frac{2}{d}-\frac{4}{d^2}+\frac{4}{d^3})C_{l_1}(\sigma).
\end{aligned}
\end{equation*}
We obtain $\frac{d-1}{d+1}<c_3$, and when $C_{l_1}(\sigma)=\frac{d-1}{d+1}$, regardless of whether it is case A or case B, $T_{\mathcal{B}}(\sigma)=\frac{2(d^3-2d+1)}{d^3}(1+C_{l_1}(\sigma))$.

\subsection{Texture of the general isotropic states}
The general isotropic states $\tau$ can be rewritten as:
\begin{equation*}\label{}
\tau=\frac{1-\vartheta}{d^2-1}\sum^d_{a,b=1}\ket{ab}\bra{ab}
+\frac{d^2\vartheta-1}{d(d^2-1)}\sum^d_{a,a'=1}\ket{aa}\bra{a'a'}.
\end{equation*}
In the computational basis $\{\ket{ab}\}^{d}_{a,b=1}$, the matrix elements are:
 \begin{equation*}\label{}
\tau_{ab,a'b'}=\frac{1-\vartheta}{d^2-1}\delta_{aa'}\delta_{bb'}
+\frac{d^2\vartheta-1}{d(d^2-1)}\delta_{ab}\delta_{a'b'}.
\end{equation*}
Denote $\overline{A}=\frac{1-\vartheta}{d^2-1}$ and $\overline{B}=\frac{d^2\vartheta-1}{d(d^2-1)}$.
Let $\overline{k}=d^2\vartheta-1\in[-1,d^2-1]$. Then
$\overline{A}=\frac{1}{d^2}-\frac{\overline{k}}{d^2(d^2-1)}$, $\overline{B}=\frac{\overline{k}}{d(d^2-1)}$.

Type I: When $a=a'$, $b=b'$ and $a=b$, \(\tau\) has $n_1=d$ diagonal elements $\tau_{aa,aa}=\overline{A}+\overline{B}$.

Type II: When $a=a'$, $b=b'$ but $a\neq b$, \(\tau\) has $n_2=d^2-d$ diagonal elements $\tau_{ab,ab}=\overline{A}$.

Type III: When $a=b$, $a'=b'$ but $a\neq a'$, \(\tau\) has  $n_3=d^2-d$ off diagonal elements $\tau_{aa,a'a'}=\overline{B}$. The remaining $n_0=d^4-2d^2+d$ matrix elements are zero.

We denote $\overline{v_1}=|\overline{A}+\overline{B}|$, $\overline{v_2}=|\overline{A}|$, and $\overline{v_3}=|\overline{B}|$. Then the $l_1$ norm coherence measure $C_{l_1}(\tau)=|n_3\overline{v_3}|=\frac{|\overline{k}|}{d+1}$. The total sum of the matrix elements modules is  $n_1|\overline{A}+\overline{B}|+n_2|\overline{A}|+n_3|\overline{B}|=1+C_{l_1}(\tau)$, and the average value is $A_{\mathcal{B}}(\tau)=\frac{1+C_{l_1}(\tau)}{d^4}$.

Next, we discuss two cases separately: $\overline{k}\geq0$ and $\overline{k}<0$.

Case A: $\overline{k}\geq0$, that is, $y\geq\frac{1}{d^2}$. In this case, $\overline{k}=d^2\vartheta-1\in[0,d^2-1]$, $C_{l_1}(\tau)=\frac{\overline{k}}{d+1}\in[0,d-1]$, $\overline{k}=(d+1)C_{l_1}(\tau)$.
Then
$\overline{A}+\overline{B}=\frac{1}{d^2}+\frac{(d-1)\overline{k}}{d^2(d^2-1)}=\frac{1+C_{l_1}(\tau)}{d^2}$,
$\overline{A}=\frac{1}{d^2}-\frac{\overline{k}}{d^2(d^2-1)}=\frac{1}{d^2}-\frac{C_{l_1}(\tau)}{d^2(d-1)}$ and $\overline{B}=\frac{\overline{k}}{d(d^2-1)}=\frac{C_{l_1}(\tau)}{d(d-1)}$, which
are all positive. Hence, $\overline{v_1}=\frac{1+C_{l_1}(\tau)}{d^2}$, $\overline{v_2}=\frac{1}{d^2}-\frac{C_{l_1}(\tau)}{d^2(d-1)}$ and $\overline{v_3}=\frac{C_{l_1}(\tau)}{d(d-1)}$.

We obtain $\overline{v_1}-A_{\mathcal{B}}(\tau)=(1+C_{l_1}(\tau))\frac{d^2-1}{d^4}>0$, $\overline{v_2}\geq A_{\mathcal{B}}(\tau)$ if and only if $C_{l_1}(\tau)\leq\frac{(d^2-1)(d-1)}{d^2+d-1}$. Denote $c_2=\frac{(d^2-1)(d-1)}{d^2+d-1}$. It follows that when $C_{l_1}(\tau)\leq c_2$, $\overline{v_2}\geq A_{\mathcal{B}}(\tau)$; conversely, when $C_{l_1}(\tau)>c_2$, $\overline{v_2}<A_{\mathcal{B}}(\tau)$.
Similarly, $\overline{v_3}\geq A_{\mathcal{B}}(\tau)$ if and only if $C_{l_1}(\tau)\geq\frac{d-1}{d^3-d+1}$. Denote $c_1=\frac{d-1}{d^3-d+1}$. Accordingly, when $C_{l_1}(\tau)\geq c_1$, then $\overline{v_3}\geq A_{\mathcal{B}}(\tau)$, whereas when $C_{l_1}(\tau)<c_1$, then $\overline{v_3}<A_{\mathcal{B}}(\tau)$.

 It can be derived that $c_1<c_2<d-1$. Therefore, under Case A, the range of $C_{l_1}(\tau)$ can be divided into the following three intervals:

Interval $\mathrm{Ai}$, $0\leq C_{l_1}(\tau)< c_1$: $\overline{v_1}>A_{\mathcal{B}}(\tau)$, $\overline{v_2}>A_{\mathcal{B}}(\tau)$ and $\overline{v_3}<A_{\mathcal{B}}(\tau)$;

Interval $\mathrm{Aii}$, $c_1\leq C_{l_1}(\tau)\leq c_2$: $\overline{v_1}>A_{\mathcal{B}}(\tau)$, $\overline{v_2}\geq A_{\mathcal{B}}(\tau)$ and $\overline{v_3}\geq A_{\mathcal{B}}(\tau)$;

Interval $\mathrm{Aiii}$, $c_2< C_{l_1}(\tau)\leq d-1$: $\overline{v_1}>A_{\mathcal{B}}(\tau)$, $\overline{v_2}< A_{\mathcal{B}}(\tau)$ and $\overline{v_3}> A_{\mathcal{B}}(\tau)$.

According to equation (\ref{SS}), for Interval $\mathrm{Ai}$, \\ $T_{\mathcal{B}}(\tau)=2(n_0+n_3)A_{\mathcal{B}}(\tau)-2n_3\overline{v_3}=\frac{2(d^2-1)}{d^2}-\frac{2C_{l_1}(\tau)}{d^2}$;

for Interval $\mathrm{Aii}$,\\
$T_{\mathcal{B}}(\tau)=2n_0A_{\mathcal{B}}(\tau)=\frac{2(d^3-2d+1)}{d^3}(1+C_{l_1}(\tau))$;

and for Interval $\mathrm{Aiii}$, \\
$T_{\mathcal{B}}(\tau)=2(n_2+n_0)A_{\mathcal{B}}(\tau)-2n_2\overline{v_2}=\frac{2(d-1)}{d^2}+\frac{2(d^2+d-1)C_{l_1}(\tau)}{d^2}$.

Case B: $\overline{k}<0$, that is, $\vartheta<\frac{1}{d^2}$. In this case, $\overline{k}=d^2\vartheta-1\in[-1,0)$, $C_{l_1}(\tau)=\frac{-\overline{k}}{d+1}\in(0,\frac{1}{d+1}]$, $\overline{k}=-(d+1)C_{l_1}(\tau)$. Then $\overline{A}+\overline{B}=\frac{1}{d^2}+\frac{(d-1)\overline{k}}{d^2(d^2-1)}=\frac{1-C_{l_1}(\tau)}{d^2}>0$,
$\overline{A}=\frac{d^2-k-1}{d^2(d^2-1)}=\frac{1}{d^2}+\frac{C_{l_1}(\tau)}{d^2(d-1)}$ and $\overline{B}=\frac{k}{d(d^2-1)}=-\frac{C_{l_1}(\tau)}{d(d-1)}$. That is,
$\overline{v_1}=\frac{1-C_{l_1}(\tau)}{d^2}$, $\overline{v_2}=\frac{1}{d^2}+\frac{C_{l_1}(\tau)}{d^2(d-1)}$ and $\overline{v_3}=\frac{C_{l_1}(\tau)}{d(d-1)}$.

Through calculations, we obtain  $\overline{v_1}\geq A_{\mathcal{B}}(\tau)$ if and only if $C_{l_1}(\tau)\leq\frac{d^2-1}{d^2+1}$. Denote $c_3=\frac{d^2-1}{d^2+1}$. It follows that when $C_{l_1}(\tau)\leq c_3$, then $\overline{v_1}\geq A_{\mathcal{B}}(\tau)$; conversely, when $C_{l_1}(\tau)>c_3$, then $\overline{v_1}<A_{\mathcal{B}}(\tau)$. $\overline{v_2}-A_{\mathcal{B}}(\tau)=\frac{d^2-1}{d^4}+\frac{d^2-d+1}{d^4(d-1)}C_{l_1}(\tau)>0$. As for $v_3$, it is the same as Case A.

It can be derived that $c_1<\frac{1}{d+1}<c_3$. Therefore, under Case B, the range of $C_{l_1}(\tau)$ can be divided into the following two intervals:

Interval BI, $0\leq C_{l_1}(\tau)< c_1$: $\overline{v_1}> A_{\mathcal{B}}(\tau)$, $\overline{v_2}>A_{\mathcal{B}}(\tau)$ and $\overline{v_3}<A_{\mathcal{B}}(\tau)$;

Interval BII, $c_1\leq C_{l_1}(\tau)\leq \frac{1}{d+1}$: $\overline{v_1}> A_{\mathcal{B}}(\tau)$, $\overline{v_2}>A_{\mathcal{B}}(\tau)$ and $\overline{v_3}\geq A_{\mathcal{B}}(\tau)$;

According to equation (\ref{SS}), for Interval BI and Interval BII, $T_{\mathcal{B}}(\tau)$ is the same as intervals $\mathrm{Ai}$ and $\mathrm{Aii}$ in Case A, respectively. We obtain $\frac{1}{d+1}<c_2$ and when $C_{l_1}(\tau)=\frac{1}{d+1}$, regardless of whether it is Case A or Case B, $T_{\mathcal{B}}(\tau)=2(1+C_{l_1}(\tau))\frac{d^3-2d+1}{d^3}$.

\subsection{Proof of Theorem 2}

\begin{proof}
Applying the Cauchy-Schwarz inequality, we obtain
\begin{equation}\label{T-CS-bound}
\begin{aligned}
T_{\mathcal{B}}(\rho)=&\sum^d_{i,j=1}\left||\rho_{ij}|-A_{\mathcal{B}}(\rho)\right|
\leq d\sqrt{\sum_{i,j}(|\rho_{ij}|-A_{\mathcal{B}}(\rho))^2}\\
=&d\sqrt{P(\rho)-\frac{(C_{l_1}(\rho)+1)^2}{d^2}}\\
=&\sqrt{d^2P(\rho)-(C_{l_1}(\rho)+1)^2}.
\end{aligned}
\end{equation}
Therefore, we get
\begin{equation*}\label{CT1}
\begin{aligned}
T_{\mathcal{B}}(\rho)+C_{l_1}(\rho)&\leq\sqrt{d^2P(\rho)-(C_{l_1}(\rho)+1)^2}+C_{l_1}(\rho)\\
&\equiv f(C_{l_1}(\rho)).
\end{aligned}
\end{equation*}
Maximizing $f(C_{l_1}(\rho))$ yields $C_{l_1}(\rho)=d\sqrt{\frac{P(\rho)}{2}}-1$, leading to
\begin{equation*}
T(\rho)+C_{l_1}(\rho) \leq d\sqrt{2P(\rho)}-1.
\end{equation*}
However, the saturation of (\ref{T-CS-bound}) requires
\begin{equation} \label{CS-equality-condition}
\left||\rho_{ij}|-A_{\mathcal{B}}(\rho)\right|=r \quad \text{for all } i,j,
\end{equation}
for some constant $r$.  Under $C_{l_1}(\rho)=d\sqrt{\frac{P(\rho)}{2}}-1$ and $\sum_{i,j}
(|\rho_{ij}|-A(\rho))^2=P(\rho)-\frac{(C_{l_1}(\rho)+1)^2}{d^2}=d^2r^2$,
we have $A_{\mathcal{B}}(\rho)=r=\frac{1}{d}\sqrt{\frac{P(\rho)}{2}}$, and (\ref{CS-equality-condition}) implies
\begin{equation*}
|\rho_{ij}| \in \left\{0, 2A_{\mathcal{B}}(\rho)\right\} = \left\{0, c\right\}, \quad c \equiv \frac{\sqrt{2P(\rho)}}{d}.
\end{equation*}
Let $N$ be the number of nonzero matrix elements. Since $C_{l_1}(\rho)+1=Nc$, we obtain $N=\frac{d^2}{2}$, so $d$ must be even.

For diagonal elements, $\rho_{ii}\in\{0,c\}$. If $m$ is the number of nonzero diagonal elements, $\mathrm{tr}\rho=1$ gives $mc=1$. Hence,
\begin{equation*}
m=\frac{d}{\sqrt{2P(\rho)}}.
\end{equation*}
Since $m\leq d$, saturation requires $P(\rho)\geq \frac{1}{2}$.

The positivity $\rho\geq0$ guarantees $|\rho_{ij}|^2 \leq \rho_{ii}\rho_{jj}$. If $\rho_{ii}=0$, then $\rho_{ij}=0$ for all $j$. Conversely, if $\rho_{ij}\neq0$, then $|\rho_{ij}|=c$ and $\rho_{ii}=\rho_{jj}=c$, which saturates the positive semidefinite condition:
\begin{equation}\label{PSD-saturation}
|\rho_{ij}|^2 = \rho_{ii}\rho_{jj}.
\end{equation}
By employing the Gram representation $\rho_{ij}=\langle v_i|v_j\rangle$ and $\rho_{ii}=\|v_i\|^2$, equation (\ref{PSD-saturation}) is transformed into $|\langle v_i|v_j\rangle| = \|v_i\|\|v_j\|$. Since this saturates the Cauchy-Schwarz inequality, the vectors $|v_i\rangle$ and $|v_j\rangle$ are collinear.

Hence, the nonzero diagonal indices split into disjoint sets  $S_1,\dots,S_q$ of collinear Gram vectors, with sizes $r_\alpha=|S_\alpha|$. Up to a basis permutation, each principal block is rank one:
\begin{equation*}
\rho^{(\alpha)}_{ij} = c\,e^{\sqrt{-1}(\theta_i-\theta_j)}, \quad i,j\in S_\alpha.
\end{equation*}
Define $|\psi_\alpha\rangle = \frac{1}{\sqrt{r_\alpha}}\sum_{j\in S_\alpha}e^{\sqrt{-1}\theta_j}|j\rangle$. Using $c=\frac{1}{m}$, we have
\begin{equation*}
\rho = \sum_{\alpha=1}^{q} \frac{r_\alpha}{m} |\psi_\alpha\rangle \langle\psi_\alpha|,
\end{equation*}
where $|\psi_\alpha\rangle$ have mutually orthogonal supports. Since the block sizes must satisfy the integer compatibility conditions,
\begin{equation}\label{block-sum}
\sum_{\alpha=1}^{q}r_\alpha = m = \frac{d}{\sqrt{2P(\rho)}},~~~ \sum_{\alpha=1}^{q}r_\alpha^2 = \frac{d^2}{2},
\end{equation}
the bound is not attainable for arbitrary $d$ and $P(\rho)$, but only when equation (\ref{block-sum}) is satisfied.
\end{proof}

%

\subsection{Proof of Theorem 3}

Before proving Theorem 3, we first give the following lemma.

\begin{lemma} \label{L1}
For any quantum states $\rho$ and $\varsigma$, the absolute value of the difference between $T_{\mathcal{B}}(\rho)$ and $T_{\mathcal{B}}(\varsigma)$ satisfies the following inequality,
\begin{equation}
|T_{\mathcal{B}}(\rho)-T_{\mathcal{B}}(\varsigma)|\leq2\sum_{ij}|\rho_{ij}-\varsigma_{ij}|,
\end{equation}
where $\rho_{ij}$ and $\varsigma_{ij}$ are the entries of $\rho$ and $\varsigma$, respectively.
\end{lemma}
Equivalently, $T_B$ is globally Lipschitz continuous with respect to the entrywise $\ell_1$ distance
$d_{1,\mathcal{B}}(\rho,\varsigma):=\sum_{i,j}|\rho_{ij}-\varsigma_{ij}|$
in the fixed reference basis $\mathcal{B}$.

\begin{proof}
Let $M_{ij}=|\rho_{ij}|$
and $A_{\mathcal{B}}(\rho)=\frac{\sum_{kl}M_{kl}}{d^2}$. Similarly, define $M'_{ij}=|\varsigma_{ij}|$ and $A_{\mathcal{B}}(\varsigma)=\frac{\sum_{kl}M'_{kl}}{d^2}$. Then $T_{\mathcal{B}}(\rho)=\sum_{ij}|M_{ij}-A_{\mathcal{B}}(\rho)|$ and $T_{\mathcal{B}}(\varsigma)=\sum_{ij}|M'_{ij}-A_{\mathcal{B}}(\varsigma)|$.
By utilizing the triangle inequality, we obtain
\begin{equation*}\label{}
\begin{aligned}
|T_{\mathcal{B}}(\rho)-T_{\mathcal{B}}(\varsigma)|&=\left|\sum_{ij}|M_{ij}-A_{\mathcal{B}}(\rho)|-\sum_{ij}|M'_{ij}-A_{\mathcal{B}}(\varsigma)|\right|\\
&\leq\sum_{ij}||M_{ij}-A_{\mathcal{B}}(\rho)|-|M'_{ij}-A_{\mathcal{B}}(\varsigma)||  \\
&\leq\sum_{ij}|(M_{ij}-A_{\mathcal{B}}(\rho))-(M'_{ij}-A_{\mathcal{B}}(\varsigma))|\\
&=\sum_{ij}|(M_{ij}-M'_{ij})-(A_{\mathcal{B}}(\rho)-A_{\mathcal{B}}(\varsigma))|,
\end{aligned}
\end{equation*}
where the last inequality originates from triangle inequality:
\begin{equation}\label{iq2}
||a_{k}|-|b_{k}||\leq|a_k-b_k|,
\end{equation}
where \(\{a_k\}_{k=1}^d\) and \(\{b_k\}_{k=1}^d\) are both complex sequences.
By utilizing the identity equation: $A_{\mathcal{B}}(\rho)-A_{\mathcal{B}}(\varsigma)=\frac{1}{d^2}\sum_{kl}(M_{kl}-M'_{kl})$,
we obtain
\begin{equation*}
\begin{aligned}\label{}
|T_{\mathcal{B}}(\rho)-T_{\mathcal{B}}(\varsigma)|&\leq\sum_{ij}|(M_{ij}-M'_{ij})-(A_{\mathcal{B}}(\rho)
-A_{\mathcal{B}}(\varsigma))|\\
&=\sum_{ij}|(M_{ij}-M'_{ij})-\frac{1}{d^2}\sum_{kl}(M_{kl}-M'_{kl})|.
\end{aligned}
\end{equation*}
By applying the triangle inequality again, we obtain
\begin{equation*}\label{}
\begin{aligned}
&|(M_{ij}-M'_{ij})-\frac{1}{d^2}\sum_{kl}(M_{kl}-M'_{kl})|,\\
&\leq|M_{ij}-M'_{ij}|+\frac{1}{d^2}\sum_{kl}|M_{kl}-M'_{kl}|.
\end{aligned}
\end{equation*}
Then
\begin{equation*}
\begin{aligned}\label{}
&\sum_{ij}|M_{ij}-M'_{ij}|+\frac{1}{d^2}\sum_{ij}\sum_{kl}|M_{kl}-M'_{kl}|\\
&=\sum_{ij}|M_{ij}-M'_{ij}|+\sum_{kl}|M_{kl}-M'_{kl}|\\
&=2\sum_{ij}|M_{ij}-M'_{ij}|.
\end{aligned}
\end{equation*}
According to inequality (\ref{iq2}), we have
$|M_{ij}-M'_{ij}|=||\rho_{ij}|-|\varsigma_{ij}||\leq|\rho_{ij}-\varsigma_{ij}|$. Therefore, the proof is completed.
\end{proof}


We now give the proof of Theorem 3.

{\it Proof}.
Since $t\mapsto\rho_t$ is absolutely continuous on $[0,T]$, every matrix element $\rho_{ij}(t)$ is absolutely continuous and satisfies
\begin{equation}
(\rho_T)_{ij}-(\rho_0)_{ij}
=\int_0^T\frac{d}{dt}\rho_{ij}(t)\,dt ,
\label{eq:rho_FTC}
\end{equation}
where $\frac{d}{dt}\rho_{ij}(t)$ exists almost everywhere.

Therefore,
\begin{equation*}\label{}
|(\rho_{T})_{ij}-(\rho_{0})_{ij}|\leq\int^T_0|\frac{d}{dt}\rho_{ij}(t)|dt.
\end{equation*}
Summing over the indices $i$ and $j$ on both sides yields.
\begin{equation*}\label{}
\begin{aligned}
\sum_{ij}|(\rho_{T})_{ij}-(\rho_{0})_{ij}|
\leq& \sum_{ij}\int^T_0|\frac{d}{dt}\rho_{ij}(t)|dt\\
=&\int^T_0\sum_{ij}|\frac{d}{dt}\rho_{ij}(t)|dt.
\end{aligned}
\end{equation*}
Employing Lemma 1, we obtain
\begin{equation*}\label{}
\begin{aligned}
|T_{\mathcal{B}}(\rho_T)-T_{\mathcal{B}}(\rho_0)|
\leq& 2\sum_{ij}|(\rho_{T})_{ij}-(\rho_{0})_{ij}| \\
\leq& 2\int^T_0\sum_{ij}|\frac{d}{dt}\rho_{ij}(t)|dt\\
=&2T\,\overline{v}_{\mathcal B}.
\end{aligned}
\end{equation*}

For $\overline{v}_{\mathcal B}>0$, we have
\begin{equation*}
T\geq\frac{|T_{\mathcal B}(\rho_T)-T_{\mathcal B}(\rho_0)|}{
2\overline{v}_{\mathcal B}}
=T_{\mathrm{TSL}}.
\end{equation*}
If $\overline{v}_{\mathcal B}=0$, then
$\frac{d}{dt}\rho_{ij}(t)=0$ almost everywhere for all $i,j$. Thus, absolute continuity implies that
$\rho_t=\rho_0$ throughout $[0,T]$, so that
$T_{\mathcal B}(\rho_T)=T_{\mathcal B}(\rho_0)$.
Therefore, no nonzero texture variation occurs in this case.

Importantly,
$|\dot{\rho}_{ij}(t)|$ in the proof is the modulus of the
derivative $\dot{\rho}_{ij}(t)$, not the derivative of the modulus
$|\rho_{ij}(t)|$. In particular, even if $\rho_{ij}(t_0)=0$ at some
$t_0$, the derivative $\dot{\rho}_{ij}(t_0)$ may still exist and
$|\dot{\rho}_{ij}(t_0)|$ is then well defined, whereas
$\frac{d}{dt}|\rho_{ij}(t)|$ may fail to exist at $t_0$. The latter
derivative is never used in our proof.



\subsection{Derivation of $T_{TSL}$}
To estimate the lower bound in Theorem \ref{TH3}, we need the following quantities:
\begin{equation}\label{1A}
\begin{aligned}
T_{\mathcal{B}}(\rho_T)=&\left|ax^2-\frac{1+2|s|x}{4}\right|+\left|1-ax^2-\frac{1+2|s|x}{4}\right|\\
+&\frac{1-2|s|x}{2},
\end{aligned}
\end{equation}
\begin{equation}\label{2A}
\begin{aligned}
T_{\mathcal{B}}(\rho_0)=\left|a-\frac{1+2|s|}{4}\right|+\left|1-a-\frac{1+2|s|}{4}\right|+\frac{1-2|s|}{2},
\end{aligned}
\end{equation}
\begin{equation}\label{}
\begin{aligned}
\sum_{ij}|\dot{\rho_{ij}}(t)|=&|\dot{\rho_{00}}(t)|+|\dot{\rho_{11}}(t)|+|\dot{\rho_{01}}(t)|+|\dot{\rho_{10}}(t)|\\
=&\gamma a e^{-\frac{\gamma t}{2}}+\frac{\gamma}{2}se^{-\frac{\gamma t}{4}}
\end{aligned}
\end{equation}
and
\begin{equation}\label{1B}
\begin{aligned}
2\langle\sum_{ij}|\dot{\rho_{ij}}(t)|\rangle_{T}=&\frac{2}{T}\int^T_0\sum_{ij}|\dot{\rho_{ij}}(t)|dt\\
=&\frac{4a}{T}(1-e^{-\frac{\gamma T}{2}})+\frac{4s}{T}(1-e^{-\frac{\gamma T}{4}}),
\end{aligned}
\end{equation}
where $a=\cos^2\frac{\theta}{2}$, $s=\sin\frac{\theta}{2}\cos\frac{\theta}{2}=\frac{1}{2}\sin\theta$ and $x=e^{-\frac{\gamma T}{4}}$.

Combining equations (\ref{1A}), (\ref{2A}) and (\ref{1B}), we obtain
\begin{widetext}
\begin{align}\label{31}
T_{TSL}&=\frac{|T_{\mathcal{B}}(\rho_T)-T_{\mathcal{B}}(\rho_0)|}{2\langle\sum_{ij}|\dot{\rho_{ij}}(t)|\rangle  _{T}}   \notag
=\frac{\left|\left|ax^2-\frac{1+2|s|x}{4}\right|
+\left|1-ax^2-\frac{1+2|s|x}{4}\right|
+\frac{1-2|s|x}{2}-\left|a-\frac{1+2|s|}{4}\right|
-\left|1-a-\frac{1+2|s|}{4}\right|
-\frac{1-2|s|}{2}\right|
}{\frac{4a}{T}(1-e^{-\frac{\gamma T}{2}})+\frac{4s}{T}(1-e^{-\frac{\gamma T}{4}})}
\\   \notag
=&\frac{|\left|\cos^2\frac{\theta}{2}e^{-\frac{\gamma T}{2}}-\frac{1+2|\frac{\sin\theta}{2}|e^{-\frac{\gamma T}{4}}}{4}\right|
+\left|1-\cos^2\frac{\theta}{2}e^{-\frac{\gamma T}{2}}-\frac{1+2|\frac{\sin\theta}{2}|e^{-\frac{\gamma T}{4}}}{4}\right|
+\frac{1-2|\frac{\sin\theta}{2}|e^{-\frac{\gamma T}{4}}}{2}
}{\frac{4\cos^2\frac{\theta}{2}}{T}(1-e^{-\frac{\gamma T}{2}})+\frac{2\sin\theta}{T}(1-e^{-\frac{\gamma T}{4}})}\\
-&\frac{
\left|\cos^2\frac{\theta}{2}-\frac{1+2|\frac{\sin\theta}{2}|}{4}\right|
+\left|1-\cos^2\frac{\theta}{2}-\frac{1+2|\frac{\sin\theta}{2}|}{4}\right|
+\frac{1-2|\frac{\sin\theta}{2}|}{2}|}{\frac{4\cos^2\frac{\theta}{2}}{T}(1-e^{-\frac{\gamma T}{2}})+\frac{2\sin\theta}{T}(1-e^{-\frac{\gamma T}{4}})}.
\end{align}
\end{widetext}

\begin{acknowledgments}
M.J. Zhao thanks the center for Quantum Information, Institute for
Interdisciplinary Information Sciences, Tsinghua University for hospitality. This work was supported by National Natural Science Foundation of
China (Grant Nos. 12171044, 12175147, 12526648); the Natural Science Foundation of Hunan Province (Grant No. 2025JJ60025); the Scientific Research Project of the Education Department of Hunan Province (Grant No. 24B0298); the specific research fund of
the Innovation Platform for Academicians of Hainan Province;
the ``R\&D Program of Beijing Municipal Education Commission (KM 202310011012)".
\end{acknowledgments}

\bibliography{2026-aqt-xc}

\end{document}